\documentclass[12pt]{article}

\usepackage{amsmath,amssymb,graphicx,multicol} 
\usepackage{epsf,color}
\usepackage{pstricks}
\usepackage{cite}
\usepackage{verbatim}

\newcommand{\beq}{\begin{eqnarray}}
\newcommand{\eeq}{\end{eqnarray}}

\newcommand{\centeron}[2]{{\setbox0=\hbox{#1}\setbox1=\hbox{#2}\ifdim
                                        
\wd1>\wd0\kern.5\wd1\kern-.5\wd0\fi
\copy0

\kern-.5\wd0\kern-.5\wd1\copy1\ifdim\wd0>\wd1
                                       \kern.5\wd0\kern-.5\wd1\fi}}
\newcommand{\ltap}{\>\centeron{\raise.35ex\hbox{$<$}}
                               {\lower.65ex\hbox{$\sim$}}\>}
\newcommand{\gtap}{\>\centeron{\raise.35ex\hbox{$>$}}
                               {\lower.65ex\hbox{$\sim$}}\>}

\newcommand\ZZ{\hbox{\zfont Z\kern-.4emZ}}
\font\zfont = cmss10 

\begin{document}
\begin{titlepage}

\vskip.5cm
\begin{center}
{\huge \bf 
The Unhiggs}

\vskip.1cm
\end{center}
\vskip0.2cm

\begin{center}
{\bf
{David Stancato}, {\rm and}
{John Terning}}
\end{center}
\vskip 8pt

\begin{center}
{\it
Department of Physics, University of California, Davis, CA  
95616} \\
\vspace*{.5cm}
\vspace*{0.3cm}
{\tt  dastancato@ucdavis.edu, jterning@gmail.com}
\end{center}

\vglue 0.3truecm

\begin{abstract}
\vskip 3pt
\noindent
We examine a scenario where the Higgs is part of an approximate conformal field theory, and has a scaling dimension greater than one.  Such an unparticle Higgs (or Unhiggs) can still break electroweak symmetry and unitarize $WW$ scattering, but its gauge couplings are suppressed. An Unhiggs model has a reduced sensitivity of the weak scale to the cutoff, and can thus provide a solution to the little hierarchy problem.

\end{abstract}

\end{titlepage}

\newpage


\setcounter{footnote}{0}
\section{Introduction}
\label{sec:intro}
\setcounter{equation}{0}

Recently Georgi \cite{Georgi,Georgi2} has introduced a new way of studying conformal sectors that couple to the standard model using two-point functions of operators with scaling dimension between one and two.  Formally the phase space corresponding to the spectral density of this two-point function resembles the phase space for a fractional number of particles, hence the name ``unparticles."
 In the context of electroweak symmetry breaking these ideas have been applied in models where the Higgs couples to an approximately conformal sector and can mix with an unparticle \cite{Irvine,mixing,Kikuchi}.  Here however we are interested in models where the Higgs itself emerges from an approximately conformal sector, or in other words the Higgs itself is an unparticle (see ref. \cite{other} for work on related ideas). 

A model with an unparticle  Higgs (aka Unhiggs) requires that we be able to gauge the kinetic term of an unparticle effective action, but this can be done in a fairly straightforward way \cite{coloredunparticles}. We also know that new excitations that couple to weak interactions  cannot be arbitrarily light, since we would have seen such states in low-energy experiments.  So the conformal symmetry of the Higgs sector must be broken not too far below the weak scale.  Fortunately there is a simple way to account for such an threshold as well \cite{Irvine,coloredunparticles,AdSCFTUnP}.  In the limit where the scaling dimension of the unparticle approaches its canonical value, the threshold becomes an ordinary mass.
This threshold, by itself, is not enough to account for a vacuum expectation value (VEV) for the Unhiggs.  However, coupling an Unhiggs to the standard model (SM) fields will force  additional conformal symmetry breaking effects at loop level and induce a potential for the Unhiggs.  Experience with the SM and its extensions suggests that top quark loops would tend to produce the largest effects (due to the large top Yukawa coupling) and that top loops also tend to destabilize the symmetric vacuum with a vanishing VEV.  For this paper we will simply assume that a suitable effective potential can be arranged and then explore the consequences of the resulting VEV for the electroweak gauge bosons.

Effective actions for unparticles are somewhat unfamiliar since they must be non-local in position space, but this is precisely in accord with the requirements of the anti-de Sitter/ conformal field theory (AdS/CFT) correspondence \cite{AdSCFTUnP}.  It has even been checked that for fermions such non-local actions reproduce the required scaling dimension independence of anomaly factors \cite{anomalies}. These cross checks give us some confidence that unparticle actions provide a consistent effective field theory for sufficiently small scaling dimensions.

The ideas we are exploring in this paper are closely related to conformal technicolor models \cite{conformaltechnicolor} where the operator that breaks electroweak symmetry has a dimension larger than one and the square of this operator (the analogue of the Higgs mass term) is assumed to be
 larger than four.  Here we will only consider weakly coupled effective actions which restricts us to having an Unhiggs  mass operator dimension that is roughly twice as big as the Unhiggs scaling dimension.  The Unhiggs model is even more closely related to gaugephobic Higgs models \cite{gaugephobic}  where a five dimensional  AdS description is set up with a bulk Higgs that corresponds to a state with an arbitrary scaling dimension.  Taking the limit where the Higgs scaling dimension goes to infinity just gives a Randall-Sundrum model.  In the gaugephobic Higgs analysis  
only  scaling dimensions larger than two were considered, since this ensures that the hierarchy problem is solved.  Here we will be content with only addressing the little hierarchy problem (that is, why the weak scale is small compared to 10 TeV), and so we can consider scaling dimensions less than two.  This is the regime where the unparticle description is useful.  We expect that a five dimensional description would yield equivalent results, but the unparticle analysis is much easier to perform.  We expect that this will be even more of an advantage when one tries to calculate loops containing Unhiggs propagators.
 
The outline of the paper  is as follows.  We will review the inclusion of gauge interactions for an Unhiggs in the next section.  Then we will  examine $WW$ scattering and see how it is unitarized in such a model.  We will then consider the phenomenological implications for LEP bounds on the Higgs mass, where we will find that an Unhiggs can be much lighter than an SM Higgs. Next we will address the problem of the top Yukawa coupling possibly becoming non-perturbative below the cutoff scale. We will then comment on the little hierarchy problem and how an Unhiggs can reduce the sensitivity of the weak scale to the cutoff.  We will also comment on the effects of additional loop induced kinetic terms and finally present our conclusions.

\section{Gauge Interactions of the Unhiggs}
\label{sec:gaugeint}
\setcounter{equation}{0}

Consider the momentum space  effective action for an unparticle field $H$ with scaling dimension $d$ and an infrared cutoff\footnote{For a discussion of the infrared cutoff, or threshold, see  \cite{AdSCFTUnP}.}  scale $\mu$:
\beq
S_0= -\int \frac{d^4 p}{(2 \pi)^4}  \,H^\dagger  \left(-p^2+\mu^2\right)^{2-d}
H~.
 \label{action0}
\eeq
The field $H$ thus has an unparticle propagator with a threshold at $\mu$:
\beq
\Delta_H(p)= \frac{-i}{\left(-p^2+\mu^2-i\epsilon\right)^{2-d}}~,
\eeq
which approaches the usual particle propagator as $d \rightarrow 1$.
If we now include a gauge coupling\footnote{For a discussion of the spin-1 resonances  in the conformal sector with gauge boson quantum numbers, see  \cite{AdSCFTUnP}.} of this field to the standard model electroweak gauge group and a Yukawa coupling to the top quark with a cutoff scale $\Lambda$ we have
\beq
S= \int d^4 x  \,-H^\dagger  \left(D^2+\mu^2\right)^{2-d}
H- \lambda_t \,{\overline t}_R \frac{H^\dagger}{\Lambda^{d-1}} \left( \begin{array}{c} t \\ b \end{array} \right)_L + \textrm{ h.c.} ~,
 \label{action}
\eeq
where the Unhiggs field transforms under the electroweak gauge group $SU(2)_L\times U(1)_Y$ as a ${\bf 2}_{1/2}$ and $D$ is a gauge covariant derivative \cite{coloredunparticles}. The $\Lambda$ dependence in the Yukawa coupling means that $H$ is scaled so that its engineering dimension matches its scaling dimension. 
Loop corrections involving these standard model couplings will break the conformal symmetry and
give additional terms:
\beq
S_{loop}&=& \int d^4 x \,\frac{C}{\Lambda^{2d-2}} \,D_\mu H^\dagger D^\mu H-\lambda\left( \frac{H^\dagger H}{\Lambda^{2d-2}} -\frac{V^2}{2}\right)^2~.
 \label{loop-action}
\eeq
The renormalized action, $S+S_{loop}$, includes two types of masses and two types of  kinetic terms.  The loop induced potential term allows for a nontrivial VEV, while the loop induced kinetic term does not lead to any qualitatively new behavior, so we save our comments on this term for section \ref{sec:Loop}. For $d\rightarrow 1$ this model just reduces to the SM Higgs sector. 

As in the SM the instability in the potential terms forces a non-zero vacuum expectation value (VEV) for the Unhiggs:
\beq 
\langle H \rangle = \left( \begin{array}{c} 0\\  \frac{\sqrt{ \lambda V^2 \Lambda^{2d-2} -\mu^{4-2d}\Lambda^{4d-4}} }{\sqrt{2 \lambda}} \end{array}\right) \equiv  \left( \begin{array}{c} 0\\  \frac{v^d}{\sqrt{2}} \end{array}\right) ~.
\eeq
Decomposing the Unhiggs into physical and Goldstone modes we can write
\beq
H&=& \frac{1}{\sqrt{2}} e^{i T^a \pi^a/v^d}   \left( \begin{array}{c} 0\\ v^d+ h  \end{array}\right)  \\
&=&\langle H \rangle + \frac{1}{\sqrt{2}}   \left( \begin{array}{c} 0\\  h  \end{array}\right) + \Pi+\dots
\eeq
where
\beq
\Pi =   \left( \begin{array}{c}\pi^+\\ \frac{i\pi^3}{\sqrt{2}}\end{array}\right) ~.
\eeq

\begin{figure}
  \centering
  \includegraphics[height = .22\textheight]{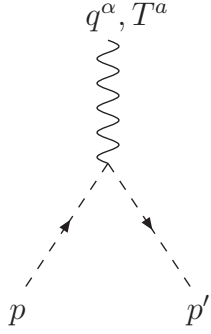}
  \caption{The one gauge boson two Unhiggs Feynman diagram (Eq. \ref{svertex})}\label{fig:1W2h}
\end{figure}

\begin{figure}
  \centering
  \includegraphics[height = .22\textheight]{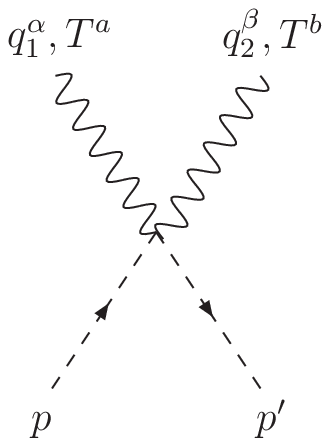}
  \caption{The two gauge boson two Unhiggs Feynman diagram (Eq. \ref{svertex2})}\label{fig:2W2h}
\end{figure}

To understand the gauge interactions it may be simpler to first write the pure derivative terms as a non-local theory in position space 
\beq
S= \int d^4x \,d^4 y \,H^\dagger(y) F(x-y)
H(x) ~,
 \label{actionx}
\eeq
and then ensure gauge invariance by using Mandelstam's method \cite{Mandelstam} of introducing 
a path-ordered exponential of the gauge field,
 i.e. a Wilson line,
 \beq
 W(x,y)=P \exp \left[-ig \tau^a \int_x^y A^a_\mu dw^\mu\right]~,
 \label{eqn:Wilson}
 \eeq
  between the two unparticle fields evaluated at $x$ and $y$ as in \cite{Mandelstam}.
  Applying this method to the electroweak interactions of the Unhiggs allows us to calculate the Feynman vertex (Figure \ref{fig:1W2h}) for a gauge boson (with gauge generator $T^a$) coupled to two Unhiggs fields.
The result  \cite{coloredunparticles} using Eq.  (\ref{action}) is 
\beq
i g \Gamma^{a\alpha}(p,q)&\equiv&\frac{i\delta^3 S}{\delta A^{a\alpha}(q)\delta \phi^\dagger(p+q)\delta \phi(p)} \nonumber\\
&=&
-i g T^a  \frac{2p^\alpha+q^\alpha}{2p \cdot q+q^2}\left[ \left(\mu^2-(p+q)^2\right)^{2-d}
-\left(\mu^2-p^2\right)^{2-d} \right]~.
\label{svertex}
\eeq
Note that this vertex satisfies the Ward-Takahashi identity \cite{wt} which relates it to the propagator $\Delta(p)$:
\beq
i q_\alpha \Gamma^{a\alpha} = \Delta^{-1}(p+q) T^a -T^a \Delta^{-1}(p)~.
\eeq
The path-ordered exponential includes arbitrarily high powers of the gauge field, so there are  vertices with arbitrary numbers of gauge bosons.  The  two gauge boson two Unhiggs vertex (Figure \ref{fig:2W2h}) is  \cite{coloredunparticles}
\beq \label{svertex2}
ig^2 \Gamma^{ab\alpha\beta}(p,q_1,q_2) &=& ig^2  \left\{ \phantom{\frac{1}{2}}\left( T^a T^b + T^b T^a \right) g^{\alpha\beta}\mathcal{F} (p,q_1+q_2)  \right. \\
&&\left.+ T^a T^b \frac{(2 p+q_2)^\beta (2 p + 2 q_2 + q_1)^\alpha}{q_1^2 + 2 (p+q_2)\cdot q_1} \left[ \mathcal{F} (p,q_1+q_2) - \mathcal{F} (p,q_2) \right]  \right.\nonumber \\
&&\left.+ T^b T^a \frac{(2 p+q_1)^\alpha (2 p + 2 q_1 + q_2)^\beta}{q_2^2 + 2 (p+q_1)\cdot q_2} \left[ \mathcal{F} (p,q_1+q_2) - \mathcal{F} (p,q_1) \right]\right\}\,,\nonumber
\eeq
where
\beq
\mathcal{F} (p,q) = -\,\frac{\left(\mu^2 - (p+q)^2\right)^{2-d}-\left(\mu^2-p^2\right)^{2-d}}{q^2 + 2 p\cdot q}\,.
\eeq
From Eq. (\ref{svertex2}) we infer that the momentum space action includes a quadratic term for gauge bosons
\beq
\label{Aquadratic}
&&- g^2A^a_\alpha A^b_\beta \langle H^\dagger \rangle T^a T^b \langle H \rangle   \\  &&\left\{ \phantom{\frac{1}{2}}g^{\alpha\beta}(d-2) \mu^{2-2d}
-  \frac{  q^\alpha q^\beta}{q^2 } \left[ (d-2)\mu^{2-2d} - \frac{\left(\mu^2 - q^2\right)^{2-d}-\left(\mu^2\right)^{2-d}}{q^2 } \right]  \right\}\,. \nonumber
\eeq
 From Eq. (\ref{svertex}) we can infer that the action also includes a gauge boson-Goldstone boson mixing term:
\beq
\mathcal{L}_{mix}= g\left( \langle H^\dagger \rangle A^a_\alpha T^a  \Pi - \Pi^\dagger  A^a_\alpha T^a \langle H \rangle\right)\,  \left[ \left(\mu^2-q^2\right)^{2-d}-\left(\mu^2\right)^{2-d}\right]q^\alpha/q^2~.
\label{Lmix}
\eeq
The last term in Eq. (\ref{Aquadratic}) is not  gauge invariant, but note that the contribution to the 
vacuum polarization from mixing with the Goldstone bosons is given by
\beq
\Pi^{ab\alpha \beta}(q)= -g^2 \langle H^\dagger \rangle  T^a
  T^b  \langle H \rangle  \,\frac{q^\alpha q^\beta}{q^4} 
\left[ \left(\mu^2-q^2\right)^{2-d}-\left(\mu^2\right)^{2-d}\right]^2 \Delta_{GB}(q).
\label{GBvacpol}
\eeq
With the Goldstone boson propagator
 \beq
\Delta_{GB}(q)&=&
-\frac{i}{\left(\mu^2 -q^2-i \epsilon \right)^{2-d} -\mu^{4-2d}}~,
   \label{GBprop1}
\eeq
the mixing contribution (\ref{GBvacpol}) cancels the gauge non-invariant term.

To discuss propagators and to perform non-trivial calculations it is convenient to remove the mixing terms by including gauge fixing terms:
\begin{align}
\mathcal{L}_{GF} = \frac{-1}{2\xi q^2 \mu^{2-2d}}\bigg[&q^{\alpha}q^{\beta}W_{\alpha}^{a}W_{\beta}^{a} + 2g\xi \mu^{2-2d}q^{\alpha}W_{\alpha}^{a}\Big(\Pi^{\dagger} T^a \langle H \rangle - \langle H^{\dagger} \rangle T^a \Pi \Big) \\ \nonumber
& -g^2\xi^2 \mu^{4-4d}\Big(\Pi^{\dagger} T^a \langle H \rangle - \langle H^{\dagger} \rangle T^a \Pi \Big)^2 \\ \nonumber
& +q^{\alpha}q^{\beta}B_{\alpha}B_{\beta} + g'\xi \mu^{2-2d}q^{\alpha}B_{\alpha}\Big(\Pi^{\dagger} \langle H \rangle - \langle H^{\dagger} \rangle \Pi \Big) \\ \nonumber
& -\frac{1}{4}g'^2\xi^2 \mu^{4-4d}\Big(\Pi^{\dagger} \langle H \rangle - \langle H^{\dagger} \rangle \Pi \Big)^2 \bigg]K(q^2)~,
\end{align}
with
\begin{align}
K(q^2) = \mu^{4-2d} - (\mu^2-q^2)^{2-d}~.
\end{align} 
Collecting all the coefficients of the quadratic terms for the $W^{\pm}$ gauge bosons we have
\beq
S_{W^+W^-}= \int\frac{d^4q}{(2 \pi)^4}\, W^{+}_\alpha(q) W^-_\beta(q) I^{\alpha \beta}
\eeq
where
\beq
I^{\alpha \beta}&=&\left(-q^2+ M_W^2\right)  g^{\alpha\beta} \\ &&
+\ \left[q^2 -M_W^2-\frac{\left(\mu^2\right)^{2-d}-\left(\mu^2-q^2\right)^{2-d}}{\mu^{2-2d}} \left(\frac{1}{\xi}-\frac{M_W^2}{(2-d)q^2}\right) \right] \frac{  q^\alpha q^\beta}{q^2 }  \,.\nonumber
\label{fullquadratic}
\eeq
and
\beq 
M_W^2 &=& \frac{g^2 (2-d)\mu^{2-2d} v^{2d}  }{4}~.
\label{eqn:Wmass}
\eeq
The propagators for the gauge bosons are then very different from in the SM:
\beq
\Delta_W(q) &=& \frac{-i}{q^2-M_W^2+i\epsilon} \\ 
&& \left(g_{\alpha\beta}+ \frac{\xi \left(q^2 -  M_W^2\right)\mu^{2-2d}-\left(\mu^{4-2d}-\left(\mu^2-q^2\right)^{2-d}\right)\left(1-\frac{\xi  \,M_W^2}{(2-d)q^2}\right)}{\left(\mu^{4-2d}-\left(\mu^2-q^2\right)^{2-d}\right)\left(q^2-\frac{\xi M_W^2}{2-d}\right)} q_\alpha q_\beta \right)~. \nonumber
\eeq
The propagators for the Goldstone bosons are then:
\beq
\Delta_{\pi^\pm}(q)&=&
-\frac{i}{\left( \mu^2-q^2-i \epsilon \right)^{2-d}-\mu^{4-2d} -\xi \frac{M_W^2}{2-d} \left[ \left(\mu^2-q^2\right)^{2-d}-\left(\mu^2\right)^{2-d}\right]/q^2}~ \\
\Delta_{\pi^3}(q)&=&
-\frac{i}{\left(\mu^2 -q^2-i \epsilon \right)^{2-d}-\mu^{4-2d} -\xi \frac{M_Z^2}{2-d} \left[ \left(\mu^2-q^2\right)^{2-d}-\left(\mu^2\right)^{2-d}\right] /q^2}~
   \label{GBprop}~,
\eeq
while for the physical Unhiggs mode we have
\beq
\Delta_{h}(q)&=&
-\frac{i}{m^{4-2d}-\mu^{4-2d}+\left( \mu^2-q^2-i \epsilon \right)^{2-d} }~
   \label{scalarprop}
\eeq
where
\beq
m^{4-2d}&=& \frac{2 \lambda v^{2d}}{ \Lambda^{4d-4}}~.
\eeq
From (\ref{scalarprop}), the location of the Unhiggs resonance is given by
\beq
M_{Unh}^2 = \mu^2 - (\mu^{4-2d} - m^{4-2d})^{\frac{1}{2-d}}~.
\label{eqn:UnHiggsmass}
\eeq
Note that the physical Unhiggs has a width if $m > \mu$ and it may also have a tachyonic mass for $m > 2 \mu$, depending on the value of $d$. To avoid these complications, we will assume $\mu > m$ in the rest of this paper.
\section{$WW$ Scattering and Unitarity}
\label{sec:production}
\setcounter{equation}{0}

The effects of unparticles on unitarity constraints have been studied in the case of $WW$ scattering \cite{UnWWscattering} and in the case of Higgs-Higgs scattering \cite{UnHHscattering}. In both of these cases, the Higgs boson was assumed to be an ordinary particle, and the unparticle belonged to a non-SM sector. The case of the Unhiggs is very different because an unparticle is replacing the SM Higgs. Thus it is important to determine whether the Unhiggs can perform the same role as the SM Higgs. One of the most important functions of the Higgs in the Standard Model is that it unitarizes longitudinal $WW$ scattering as the incoming energy becomes large. As we will show, although the Unhiggs case is more complicated and necessitates the use of a non-SM vertex, the Unhiggs is also sufficient to unitarize $WW$ scattering at high energies. 

\begin{figure}
  \centering
  \includegraphics[height = .2\textheight]{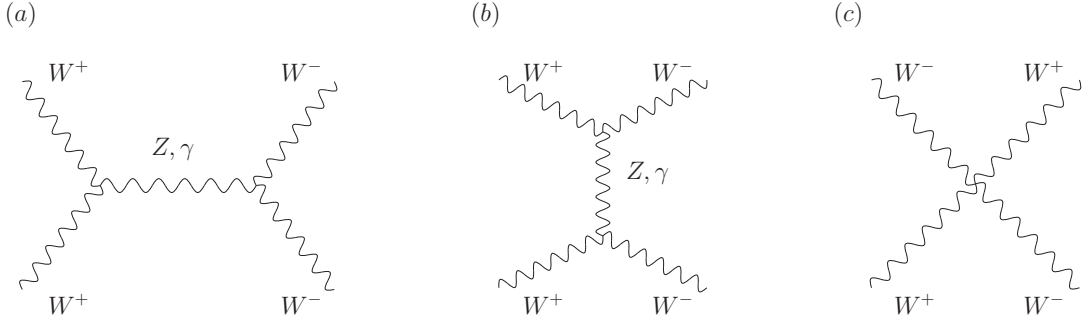}
  \caption{$WW$ scattering diagrams containing no Unhiggs vertices or propagators}\label{fig:WWSM}
\end{figure}

To calculate the $WW$ scattering diagrams, it will be easiest to use the Landau gauge, $\xi=0$,
where
\beq
\Delta_W(q)&=&\frac{-i}{q^2-M_W^2+i\epsilon} \left( 
g_{\alpha\beta}- \frac{q_\alpha q_\beta}{q^2}
\right)~,  \\
\Delta_{\pi^\pm}(q)&=&\Delta_{\pi^3}(q)=
-\frac{i}{\left( \mu^2-q^2-i \epsilon \right)^{2-d}-\mu^{4-2d} }~.
   \label{propLandau}
\eeq
Since in this gauge the gauge boson propagators are the same as in the SM Landau gauge, it is easy to see that the $WW$ scattering diagrams given in Figure \ref{fig:WWSM}, which contain no Unhiggs propagators or vertices, are the same in the Unhiggs model as in the SM. In the high energy limit, $s$, $t$ $\gg$ $M_W^2$, $M_Z^2$, the contributions of these diagrams to the $WW$ scattering amplitude are \cite{WWscattering} 
\beq
{\cal M}_{Gauge, SM} = \frac{1}{4}\frac{ig^2}{M_W^2}(s+t)~.
\label{eqn:WWSM}
\eeq
\begin{figure}
  \centering
  \includegraphics[height = .2\textheight]{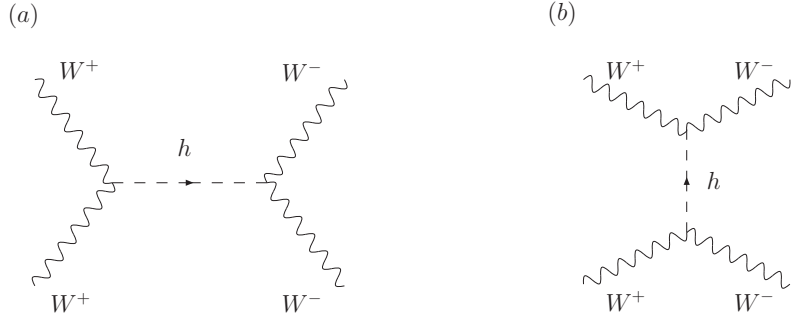}
  \caption{$WW$ scattering diagrams containing two gauge boson one Unhiggs vertices.}\label{fig:WW1UnH}
\end{figure}
The $WW$ scattering diagrams in Figure \ref{fig:WW1UnH} are analogues of  similar SM diagrams, but now contain the two gauge boson one Unhiggs vertex and an Unhiggs propagator. From (\ref{svertex2}) we see that the two gauge boson one Unhiggs vertex is given by:
\beq 
\label{eqn:2G1H}
g^2 \Gamma_1^{ab\alpha\beta}(q_1,q_2) &=& g^2  \left\{ \phantom{\frac{1}{2}}\left( T^a T^b + T^b T^a \right) \langle H\rangle g^{\alpha\beta}\mathcal{F} (0,q_1+q_2)  \right. \\
&&\left.+ T^a T^b \langle H\rangle \frac{q_2^\beta (2 q_2 + q_1)^\alpha}{q_1^2 + 2 q_2\cdot q_1} \left[ \mathcal{F} (0,q_1+q_2) - \mathcal{F} (0,q_2) \right]  \right.\nonumber \\
&&\left.+ T^b T^a \langle H\rangle \frac{q_1^\alpha (2 q_1 + q_2)^\beta}{q_2^2 + 2 q_1\cdot q_2} \left[ \mathcal{F} (0,q_1+q_2) - \mathcal{F} (0,q_1) \right]\right\}\,\nonumber\\
&& + g^2  \left\{ \phantom{\frac{1}{2}}\langle H^\dagger \rangle \left( T^a T^b + T^b T^a \right)  g^{\alpha\beta}\mathcal{F} (-q_1-q_2,q_1+q_2)  \right. \nonumber\\
&&\left.- \langle H^\dagger \rangle T^a T^b  \frac{q_1^\alpha (2 q_1 + q_2)^\beta}{q_1^2} \left[ \mathcal{F} (-q_1-q_2,q_1+q_2) - \mathcal{F} (-q_1-q_2,q_2) \right]  \right.\nonumber \\
&&\left.- \langle H^\dagger \rangle T^b T^a  \frac{q_2^\beta (2 q_2 + q_1)^\alpha}{q_2^2} \left[ \mathcal{F} (-q_1-q_2,q_1+q_2) - \mathcal{F} (-q_1-q_2,q_1) \right]\right\}\,,\nonumber
\eeq
where $q_1$ and $q_2$ are the incoming momenta of the gauge bosons. Since the polarization vectors satisfy $\epsilon (q_i)\cdot q_i =0$ and using $\mathcal{F} (0,q_1+q_2) = \mathcal{F} (-q_1-q_2,q_1+q_2)$ we  have
\beq
g^2 \epsilon_\alpha(q_1) \epsilon_\beta(q_2)\Gamma_1^{ab\alpha\beta}(q_1,q_2) &=& g^2  \left( T^a T^b + T^b T^a \right)  \langle H\rangle \, \epsilon(q_1)\cdot  \epsilon(q_2) \mathcal{F} (0,q_1+q_2) \\
&&+g^2 \langle H^\dagger \rangle \left( T^a T^b + T^b T^a \right)  \epsilon(q_1)\cdot  \epsilon(q_2) \mathcal{F} (0,q_1+q_2)~.
\eeq
Therefore, in terms of $W^{+}$, $W^{-}$ and the physical Unhiggs, we have
\beq
ig^2 \epsilon_\alpha(q_1) \epsilon_\beta(q_2) \Gamma_1^{+-\alpha\beta}(q_1,q_2) &=& i\frac{g^2 v^d}{2} \epsilon(q_1)\cdot \epsilon(q_2) \mathcal{F} (0,q_1+q_2)~.
\eeq
So for the s-channel Unhiggs exchange contribution (Figure \ref{fig:WW1UnH}b) to $WW$ scattering we find
\beq
{\cal M}_{h}(s) &=& -g^2 \frac{M_W^2}{(2-d)\mu^{2-2d} }  \left(\epsilon(q_1)\cdot  \epsilon(q_2) \frac{\left(\mu^2 - s\right)^{2-d}-\left(\mu^2\right)^{2-d}}{s} \right) \Delta_{h}(s)\\
&&\cdot  \left(\epsilon(q_3)\cdot  \epsilon(q_4) \frac{\left(\mu^2 - s\right)^{2-d}-\left(\mu^2\right)^{2-d}}{s} \right) \nonumber \\
&=& -g^2 \frac{M_W^2}{(2-d)\mu^{2-2d} }  \left( \left(1-\frac{s}{2M_W^2}\right) \frac{\left(\mu^2 - s\right)^{2-d}-\left(\mu^2\right)^{2-d}}{s} \right)^2 \Delta_{h}(s)~.
\eeq
When $s\gg M_W^2,\mu^2,m^2$ we have
\beq
{\cal M}_{h}(s)
&=&i\frac{g^2}{4M_W^2 (2-d)\mu^{2-2d}} \left( -s \right)^{2-d}~.
\label{eqn:Higgs s}
\eeq
Similarly, the t-channel Unhiggs exchange contribution (Figure \ref{fig:WW1UnH}a) to $WW$ scattering will be, for $t\gg M_W^2, \mu^2,m^2$
\beq
{\cal M}_{h}(t)
&=&i\frac{g^2}{4M_W^2 (2-d)\mu^{2-2d}} \left( -t \right)^{2-d}~.
\label{eqn:Higgs t}
\eeq
Combining the three amplitudes in (\ref{eqn:WWSM}), (\ref{eqn:Higgs s}) and (\ref{eqn:Higgs t}), we see that in the $d \rightarrow 1$ limit, the terms which grow with energy disappear, and $WW$ scattering is unitarized. However, in the case of interest, $d \neq 1$, the dangerous high energy terms do not cancel. So, unlike in the case of the SM, the Unhiggs exchange diagrams in Figure \ref{fig:WW1UnH} are insufficient to unitarize $WW$ scattering.

As noted earlier, however, the Unhiggs action allows for vertices with arbitrary numbers of gauge bosons. Therefore, there is another Unhiggs contribution to $WW$ scattering from the four gauge boson two Unhiggs vertex, given by
\beq
ig^4 \Gamma^{abcd\mu\nu\alpha\beta}(p,q_1,q_2,q_3,q_4) &\equiv& \frac{i\delta^6 S}{\delta A^{a\mu}(q_1)\delta A^{b\nu}(q_2)\delta A^{c\alpha}(q_3)\delta A^{d\beta}(q_4)\delta \phi^{\dagger} (p')\delta \phi (p)}
\eeq 
with
\beq
p' &=& p+q_1+q_2+q_3+q_4~.
\eeq
\begin{figure}
  \centering
  \includegraphics[height = .22\textheight]{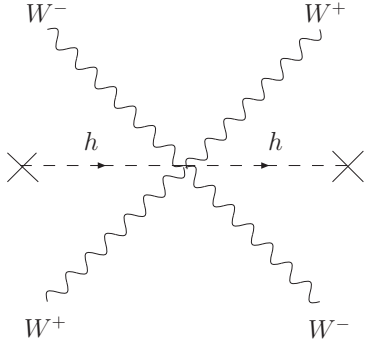}
  \caption{The four gauge boson two Unhiggs contribution to $WW$ scattering}\label{fig:WW2UnH}
\end{figure}
The four gauge boson two Unhiggs vertex which contributes to $WW$ scattering is shown in Figure \ref{fig:WW2UnH}. The crosses on the Unhiggs lines denote that we have taken $p=p'=0$ and set the Unhiggs to its VEV. Using the methods outlined in Section \ref{sec:gaugeint}, we derive an expression for the four gauge boson two Unhiggs vertex which we can then apply to $WW$ scattering.
The piece of the four gauge boson two Unhiggs vertex with no uncontracted momenta is given by
\beq
\label{eqn:4W2h}
ig^4 \Gamma^{abcd\mu\nu\alpha\beta}(p,q_1,q_2,q_3,q_4) &\ni& -ig^4 \langle H^\dagger \rangle \left\{{\cal R}^{cd} {\cal R}^{ab} g^{\mu\nu}g^{\alpha\beta} {\cal G}(p',q_1,q_2,p) \right. 
\\ 
&&\left. + {\cal R}^{bd} {\cal R}^{ac} g^{\mu\alpha}g^{\nu\beta} {\cal G}(p',q_1,q_3,p) + {\cal R}^{ac} {\cal R}^{bd} g^{\nu\beta}g^{\alpha\mu} {\cal G}(p',q_2,q_4,p) \right.\nonumber \\
&&\left. + {\cal R}^{ad} {\cal R}^{bc} g^{\nu\alpha}g^{\mu\beta} {\cal G}(p',q_2,q_3,p) + {\cal R}^{ab} {\cal R}^{cd} g^{\mu\nu}g^{\alpha\beta} {\cal G}(p',q_3,q_4,p) \right. \nonumber \\ 
&&\left.+ {\cal R}^{bc} {\cal R}^{ad} g^{\mu\beta}g^{\nu\alpha} {\cal G}(p',q_1,q_4,p) \right\}\,\langle H \rangle \nonumber 
\eeq
where 
\beq
{\cal R}^{ab} &\equiv& (T^a T^b + T^b T^a)
\eeq
and
\beq
{\cal G}(p',q_i,q_j,p) &\equiv& \left\{ \frac{\left(\mu^2 - p'^2\right)^{2-d}}{\left(p'^2-(p+q_i+q_j)^2\right)\left(p'^2-p^2\right)} \right.
\\
&&\left. + \frac{\left(\mu^2 - (p+q_i+q_j)^2\right)^{2-d}}{\left((p+q_i+q_j)^2 - p^2\right)\left((p+q_i+q_j)^2 - p'^2\right)} \right\}~. \nonumber 
\label{eqn:G}
\eeq
Taking the limit $p\rightarrow0$, $p'\rightarrow0$, we have
\beq
{\cal G}(0,q_i,q_j,0) &=& \frac{(2-d)(\mu^{2-2d})}{(q_i+q_j)^2} + \frac{\left(\mu^2-(q_i+q_j)^2\right)^{2-d}}{(q_i+q_j)^4} - \frac{\mu^{4-2d}}{(q_i+q_j)^4}~.
\eeq 

It turns out that to evaluate the $WW$ scattering amplitude arising from the four gauge boson two Unhiggs vertex we only need to consider the piece of the vertex with no uncontracted momenta, as given in (\ref{eqn:4W2h}). We can see this by looking at the Lorentz structure of the other terms in the vertex. The terms in the vertex with no $g^{\mu \nu}$ factors have a Lorentz structure given by  
\beq
(2p+q_1)^\mu (2p+2q_1+q_2)^\nu (2p+2q_2+q_3)^\alpha (2p+2q_2+2q_3+q_4)^\beta  \mbox { + permutations}.
\eeq
Upon setting $p = p' = 0$, and contracting with $\epsilon (q_1)_\mu \epsilon (q_2)_\nu \epsilon (q_3)_\alpha \epsilon (q_4)_\beta$, we find that the contribution to the amplitude is zero from this term, again using the fact that $\epsilon (q_i) \cdot q_i = 0$. 
There are also terms in the vertex which contain one factor of $g^{\mu \nu}$. These terms have a Lorentz structure given by 
\beq
&&(2p+q_1)^\mu (2p+2q_1+q_2)^\nu g^{\alpha \beta} + g^{\nu \alpha} (2p+q_1)^\mu (2p+2q_1+2q_2+2q_3+q_4)^\beta \\
&& + \; g^{\mu \nu} (2p+2q_1+2q_2+q_3)^\alpha (2p+2q_1+2q_2+2q_3+q_4)^\beta \mbox{ + permutations}. \nonumber
\label{eqn:onemetricterm}
\eeq
Upon taking $p\rightarrow 0$ and contracting with  $\epsilon (q_1)_\mu \epsilon (q_2)_\nu \epsilon (q_3)_\alpha \epsilon (q_4)_\beta$, the first two terms in (\ref{eqn:onemetricterm}) clearly go to zero. The third term in (\ref{eqn:onemetricterm}) also goes to zero upon contracting with the polarization vectors because $2p+2q_1+2q_2+2q_3+q_4 = 2p'-q_4 \rightarrow -q_4$ when taking $p' \rightarrow 0$. 

So the only contribution from the four gauge boson vertex to the $WW$ scattering amplitude comes from (\ref{eqn:4W2h}) and is given by
\beq
{\cal M}_{2h}(s) &=& \frac{-ig^4v^{2d}}{4}\left(\frac{(2-d)(\mu^{2-2d})}{s} + \frac{(\mu^2-s)^{2-d}}{s^2} - \frac{\mu^{4-2d}}{s^2}\right)\left[ \epsilon(q_1)\cdot \epsilon(q_2)\right]^2
\eeq
and
\beq
{\cal M}_{2h}(t) &=& \frac{-ig^4v^{2d}}{4}\left(\frac{(2-d)(\mu^{2-2d})}{t} + \frac{(\mu^2-t)^{2-d}}{t^2} - \frac{\mu^{4-2d}}{t^2}\right)\left[ \epsilon(q_1)\cdot \epsilon(q_3)\right]^2~.
\eeq 
Taking $s\gg M_W^2,\mu^2,m^2$ and $t\gg M_W^2, \mu^2,m^2$, we have 
\beq
\label{eqn:2Unhiggs s}
{\cal M}_{2h}(s) &=& \frac{-ig^4v^{2d}}{16M_W^4}\left[(2-d)(\mu^{2-2d})s + (-s)^{2-d}\right] \\ \nonumber
&=& -i\frac{g^2}{4M_W^2}\left[s + \frac{(-s)^{2-d}}{(2-d)(\mu^{2-2d})}\right]
\eeq
and
\beq
\label{eqn:2Unhiggs t}
{\cal M}_{2h}(t) &=& \frac{-ig^4v^{2d}}{16M_W^4}\left[(2-d)(\mu^{2-2d})t + (-t)^{2-d}\right] \\ \nonumber
&=& -i\frac{g^2}{4M_W^2}\left[t + \frac{(-t)^{2-d}}{(2-d)(\mu^{2-2d})}\right]~.
\eeq
The total $WW$ scattering amplitude is given by 
\beq
{\cal M} &=& {\cal M}_{Gauge,SM} + {\cal M}_{h} + {\cal M}_{2h}~.
\eeq 
Combining Eqs. (\ref{eqn:WWSM}), (\ref{eqn:Higgs s}), (\ref{eqn:Higgs t}), (\ref{eqn:2Unhiggs s}) and (\ref{eqn:2Unhiggs t}), we get
\beq
{\cal M} &=& \frac{1}{4} \frac{ig^2}{M_W^2} (s+t) + i \frac{g^2}{4M_W^2 (2-d)\mu^{2-2d}}(-s)^{2-d} +  i \frac{g^2}{4M_W^2 (2-d)\mu^{2-2d}}(-t)^{2-d} \\ \nonumber
&& - i \frac{g^2}{4M_W^2} \left[s + \frac{(-s)^{2-d}}{(2-d)\mu^{2-2d}} \right] - i \frac{g^2}{4M_W^2} \left[t + \frac{(-t)^{2-d}}{(2-d)\mu^{2-2d}} \right] + \mathcal{O}(s^0) + \mathcal{O}(t^0)~.
\eeq
All of the terms that scale with positive powers of energy cancel in the full amplitude, and therefore the Unhiggs does unitarize $WW$ scattering.

\begin{figure} [h]
  \centering
  \includegraphics[height = .5\textheight]{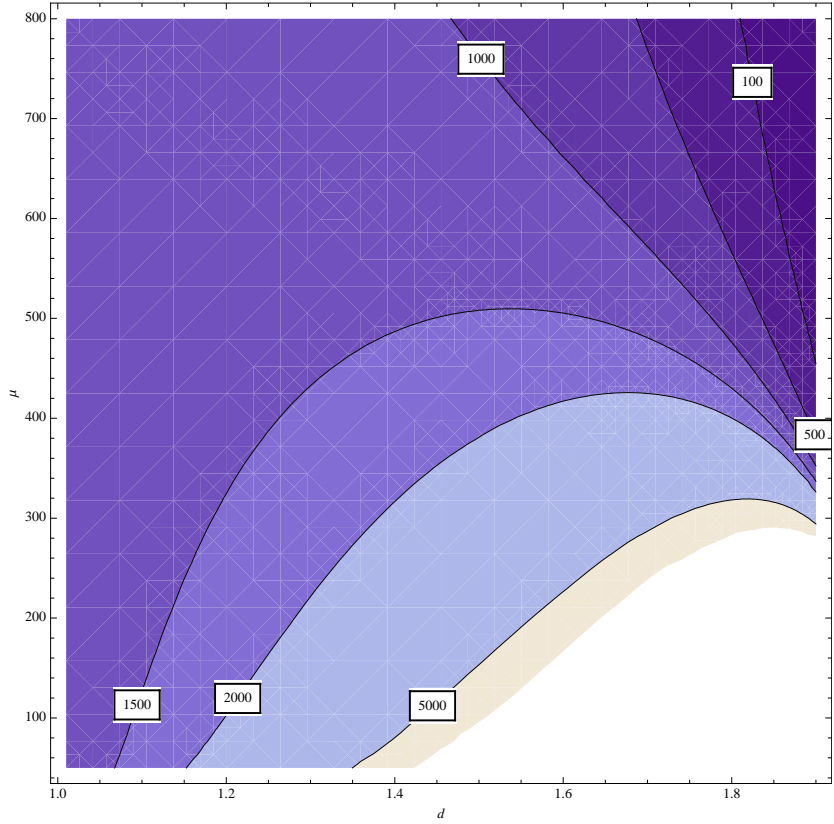}
  \caption{Contour plots  of the bound on $m$ in the $d$-$\mu$ plane. The darkest regions have the lowest upper bound on $m$, Contour lines are shown for 100, 500, 1000, 1500, 2000, and 5000 GeV.}
\label{fig:Contourplot}
\end{figure}

However, partial wave unitarity can still be violated for certain values of the parameters $\mu$, $m$, and $d$. Therefore, we now examine the  finite terms for $s, t, \mu^2, m^2 \gg M_W^2, M_Z^2$. Following \cite{HiggsHunter}, we note that there is a bound on the coefficient of the first partial wave such that 
\beq
\label{eqn:bound}
|a_0| \leq 1
\eeq
where $a_l$ is defined by the equation
\beq
{\cal M}(s,t) = 16\pi \displaystyle\sum_{l=0} (2l+1)a_l(s)P_l(\cos \theta)
\eeq
and $P_l(\cos \theta)$ is a Legendre polynomial.
Projecting out the zeroth partial wave and using $t = 2M_W^2 - \frac{s}{2}(1-\cos \theta ) \approx -\frac{s}{2}(1- \cos \theta )$, we have
\beq
\label{eqn:a0def}
a_0(s) = \frac{1}{16\pi s} \int_{-s}^0 {\cal M}(s,t) dt~.
\eeq
In the limit $s, t, \mu^2, m^2 \gg M_W^2, M_Z^2$, the full scattering amplitude is given by
\beq
{\cal M}(s,t) &=& i\frac{g^2m^{4-2d}}{4M_W^2(2-d)\mu^{2-2d}} \left\{ \frac{\mu^{4-2d} - (\mu^2-s)^{2-d}}{m^{4-2d} - \mu^{4-2d} + (\mu^2-s)^{2-d}} \right. \nonumber \\
&& \left. + \frac{\mu^{4-2d} - (\mu^2-t)^{2-d}}{m^{4-2d} - \mu^{4-2d} + (\mu^2-t)^{2-d}} \right\}~.
\eeq

Inserting ${\cal M}(s,t)$ into (\ref{eqn:a0def}) and performing the integration, we find
\beq
a_0(s) &=& i\frac{G_Fm^{4-2d}}{8\pi \sqrt{2} (2-d)\mu^{2-2d}} \left\{ \frac{\mu^{4-2d}-(\mu^2-s)^{2-d}}{m^{4-2d}-\mu^{4-2d}+(\mu^2-s)^{2-d}} \right. \nonumber \\
&& \left. -1 + \frac{1}{s}(\frac{\mu^{4-2d}}{m^{4-2d}-\mu^{4-2d}} + 1) \left[(\mu^2+s)Q_d(\mu^2+s) - \mu^2Q_d(\mu^2)\right]\right\}\,
\eeq
where $G_F = 1.166 \times 10^{-5} \textrm{ GeV}^{-2}$ is the Fermi constant and $Q_d(z)$ is defined in terms of the hypergeometric function as
\beq
Q_d(z) \equiv {}_2F_1(\frac{1}{2-d}, 1; 1+\frac{1}{2-d}; -\frac{z^{2-d}}{m^{4-2d}-\mu^{4-2d}})~.
\eeq
For large $s$ this gives the bound
\beq
G_F \, m^{4-2d} < 4 \pi \sqrt{2} (2-d) \mu^{2-2d}~.
\eeq
The resulting bound in the $d$-$\mu$ plane is shown in Figure \ref{fig:Contourplot}.
The bound is only stringent for large $d\sim 2$ and large $\mu$.

\section{Lowering the LEP bounds on the Unhiggs}
\label{sec:LEP}
\setcounter{equation}{0}

The LEP experiment put a lower bound on the Standard Model Higgs mass of $114.4 \; \textrm{GeV}$ \cite{LEP}. In the case of a non-SM Higgs, such as the Unhiggs, there are two effects which can change the lower bound. One way for the bound to be invalid is to have a branching ratio of $H \rightarrow b \overline{b}$ that is different than in the SM, eg. ref.~\cite{Dermisek:2005ar}. The other way is to suppress the cross section for Higgs production. At LEP, the Higgs is produced by the ``Higgs-strahlung'' process, $e^+ e^- \rightarrow Z^\ast \rightarrow HZ$. If the cross section for this process is suppressed relative to the SM, then the lower bound will be reduced. In the case of the Unhiggs, we expect such a suppression as $d \rightarrow 2$, because in this limit the gauge covariant derivative in Eq. (\ref{action}) disappears. The suppression is also clear from noting that the Unhiggs-gauge boson couplings in Eqs. (\ref{svertex}), (\ref{svertex2}) and (\ref{eqn:2G1H}) go to zero in the $d \rightarrow
  2$ limit. The Unhiggs becomes gaugephobic \cite{gaugephobic} when its scaling dimension is near 2  or larger \cite{AdSCFTUnP}. We can quantify the suppression with the definition
\beq
\label{eqn:Xi^2}
\xi^2 \equiv \frac{\sigma_{Unh} (e^+ e^- \rightarrow HZ)}{\sigma_{SM} (e^+ e^- \rightarrow HZ)}~.
\eeq

\begin{figure}[h]
  \centering
  \includegraphics[height = .3\textheight]{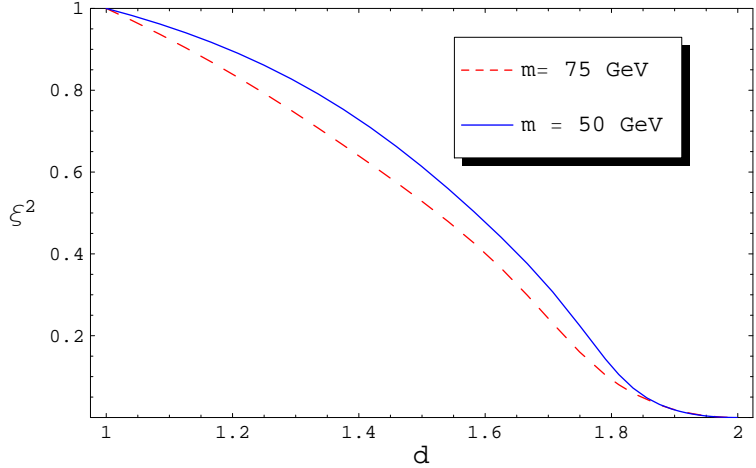}
  \caption{$\xi^2$ as a function of $d$ for $\mu = 100$ GeV, $m = 75$ GeV and $\mu = 100$ GeV, $m = 50$ GeV}
\label{fig:Xi2}
\end{figure}

We find that, upon ignoring terms proportional to the electron mass, there is a simple relation between the amplitude squared in the Unhiggs model and the amplitude squared in the Standard Model.
\beq
\label{eqn:MUnH}
|{\cal M}_{Unh} (e^+ e^- \rightarrow HZ)|^2 = \frac{|{\cal P}|^2}{(2-d) \mu^{2-2d}} |{\cal M}_{SM} (e^+ e^- \rightarrow HZ) |^2
\eeq
where 
\beq
{\cal P} = \frac{(\mu^2)^{2-d} - \left( \mu^2 - p_h^2 \right)^{2-d}}{ p_h^2}~, 
\eeq
$p_h$ is the momentum 4-vector of the outgoing Unhiggs and
\beq
|{\cal M}_{SM} (e^+ e^- \rightarrow HZ) |^2 \propto \left[ 1+\frac{E_Z^2}{M_Z^2} + (1 - \frac{E_Z^2}{M_Z^2}) \cos ^2\theta \right]~.
\eeq
 The cross section for $2 \rightarrow 2$ scattering is given in general by 
 \beq
\label{eqn:crosssection}
  \sigma = \int \frac{|{\cal M}|^2}{2s} d\Phi
 \eeq 
where $d\Phi$ is the phase space factor associated with the outgoing particles.

The phase space factor associated with the Unhiggs contains both a continuum above the IR cutoff $\mu$, and a pole at $q^2 = M_{Unh}^2$, where $M_{Unh}$ is given by Eq. (\ref{eqn:UnHiggsmass}). If $m<\mu$, then the pole is necessarily below the IR cutoff and the phase space takes a relatively simple form and is given by
\beq \nonumber
d\Phi_h(q^2) &=& \frac{A_d \theta(q^0)\theta(q^2 - \mu^2) (q^2-\mu^2)^{2-d}}{(\mu^{4-2d}-m^{4-2d})^2 + (q^2-\mu^2)^{4-2d} - 2(\mu^{4-2d}-m^{4-2d})(q^2-\mu^2)^{2-d} \cos (d\pi)} \\
&+& \theta(q^0)\frac{-\pi A_d}{\sin (d\pi)}\frac{(\mu^{4-2d}-m^{4-2d})^{\frac{d-1}{2-d}}}{(2-d)}\delta(q^2 - M_{Unh}^2)~,
\label{eqn:Phase}
\eeq
where $A_d$ is a normalization factor as in \cite{Georgi}. 
The first line in Eq. (\ref{eqn:Phase}) contains the continuum while the second line contains the pole. Note that for $\mu > m$, the continuum part of the phase space goes to zero as $d \rightarrow 1$ because of the fact that $A_{d=1} = 0$. 
Also, the part of the phase space containing the pole  has the following $d \rightarrow 1$ limit:
\beq
\displaystyle\lim_{d \to 1}d\Phi_{h,pole}(q^2) = 2\pi \theta(q^0)\delta(q^2 - m^2)~.
\eeq
Thus, for $\mu > m$, the Unhiggs phase space in Eq. (\ref{eqn:Phase}) does indeed reduce to the Standard Model Higgs result.

Using Eqs. (\ref{eqn:MUnH}), (\ref{eqn:crosssection}) and (\ref{eqn:Phase}), $\xi^2$ can be calculated numerically as a function of $d$ for any values of the parameters $\mu$ and $m$. A plot of $\xi^2$ vs. $d$ for two pairs of $\mu$ and $m$ is shown in Figure (\ref{fig:Xi2}).
As expected, $\xi^2$ falls as $d$ gets larger and is approximately zero for $d \rightarrow 2$. This shows that for moderate to high values of $d$, the suppression of the Unhiggs-Gauge couplings allows for an Unhiggs lighter than 114 GeV to have evaded detection at LEP.

\section{Running of the top Yukawa coupling}
\label{sec:TopYuk}
\setcounter{equation}{0}
Consider the top Yukawa coupling given in Eq. (\ref{action}) which  leads to an $ht \overline{t}$ interaction term 
\beq
{\cal L}_{Y} = \frac{1}{\sqrt{2}} \frac{\lambda_t}{\Lambda^{d-1}} h t \overline{t}~.
\eeq
After expanding around the Unhiggs VEV, the top Yukawa coupling is given at tree level by
\begin{align}
\frac{\lambda_t}{\Lambda^{d-1}}\frac{v^d}{\sqrt{2}} = m_t
\end{align}
which, after writing $v^d$ in terms of $M_W$ leads to
\begin{align}
\lambda_t = \frac{\sqrt{2}}{2}\frac{m_t}{M_W}g\sqrt{2-d}\left(\frac{\Lambda}{\mu}\right)^{d-1} ~.
\end{align}
In the Standard Model, $\lambda_t =  \frac{\sqrt{2}}{2}\frac{m_t}{M_W}g \approx 1$ at the electroweak scale. Therefore, with a cutoff $\Lambda = 10$ TeV, $\lambda_t$ may be significantly greater than one at the electroweak scale, depending on the values of $\mu $ and $d$. In addition, we know that $\lambda_t$ grows in the UV due to quantum corrections associated with the top Yukawa coupling. Thus we need to calculate the running of $\lambda_t$ to make sure that it does not become non-perturbative before the cutoff at around 10 TeV. We will start by defining the value of $\lambda_t$ at the electroweak scale $s_0 \approx$ 100 GeV where $M_W$ and $m_t$ are measured, so that 
\beq
\lambda_t (s_0) = \frac{\sqrt{2}}{2}\frac{m_t}{M_W}g\sqrt{2-d}\left(\frac{\Lambda}{\mu}\right)^{d-1} ~.
\eeq

To explain how we compute the running of $\lambda_t$, we start by noting that in the SM, the top Yukawa beta function gets a positive contribution from terms proportional to $\lambda_t^3$ and a negative contribution from terms proportional to the gauge couplings. Since the $U(1)_Y$ and $SU(2)_L$ gauge couplings are small compared to the top Yukawa coupling and the QCD coupling, we will ignore diagrams proportional to $g$ and $g'$. The gluon contribution will be the same in the Unhiggs model as in the SM, so it will not be necessary to perform that calculation in this paper. However, we must calculate the contribution from diagrams proportional to $\lambda_t^3$, as these will in general be different than in the SM. Considering only diagrams proportional to $\lambda_t^3$, the correction to the top quark propagator as well as the correction to the proper Yukawa vertex both contain Unhiggs propagators in the loop, whereas the correction to the Unhiggs propagator is due to a top loop and thus does not contain an unparticle propagator in the loop. This is important, due to the fact that since the unparticle propagator has a reduced power of $p^2$ compared to a regular particle propagator, any loop which does not contain an unparticle propagator will generate a subleading Unhiggs kinetic term\footnote{These kinetic terms are further discussed in Section 7.} compared to a loop with an unparticle propagator. For this reason, the correction to the Unhiggs propagator is subleading and we have thus only included the correction to the top quark propagator and the correction to the proper Yukawa vertex.    

\begin{figure}
  \centering
  \includegraphics[height = .2\textheight]{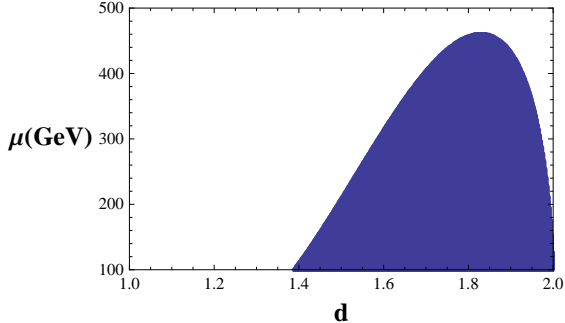}
  \caption{The white region is the region in $\mu$-$d$ parameter space where $\lambda_t$ remains perturbative up to 10 TeV, while the dark region is the region where $\lambda_t$ is non-perturbative at or below 10 TeV.}
\label{fig:TopYukCorr}
\end{figure}

Using the procedure described above and ignoring the gluon contribution for now, we find that the top Yukawa coupling at an energy scale $s$ is given by
\begin{align}
\lambda_t(s) = \lambda_t(s_0)\left[1-\frac{3\left(\lambda_t(s_0)\right)^2}{64\pi^2}\frac{1}{d-1}\left(1-\left(\frac{s_0^2}{s^2}\right)^{d-1}\right)\right]^{-1}~.
\end{align} 
The contour plot shown in Figure \ref{fig:TopYukCorr} represents the value of the top Yukawa coupling at the cutoff scale $\Lambda = $ 10 TeV as a function of $\mu$ and $d$. We require $\lambda_t(\Lambda) \leq 2\pi$ so that the coupling remains perturbative up to the cutoff. The white region of the plot is the region for which $\lambda_t(\Lambda)$ does indeed remain less than $2\pi$. It is important to note that since we did not include the gluon loops, which, as described above, causes $\lambda_t$ to decrease in the UV, the plot is a very conservative estimate of the allowed values of $\mu$ and $d$.

From Figure \ref{fig:TopYukCorr}, we see that these conservative estimates require that to have moderate to high values of $d$, we must choose $\mu$ to be somewhat above the electroweak scale. For instance, with a value of $d=1.7$, we need to choose $\mu \gtrsim 400$ GeV.

\section{Yukawa Couplings and the UV cutoff}
\label{sec:Yukawa}
\setcounter{equation}{0}

\begin{figure} [h]
  \centering
  \includegraphics[height = .25\textheight]{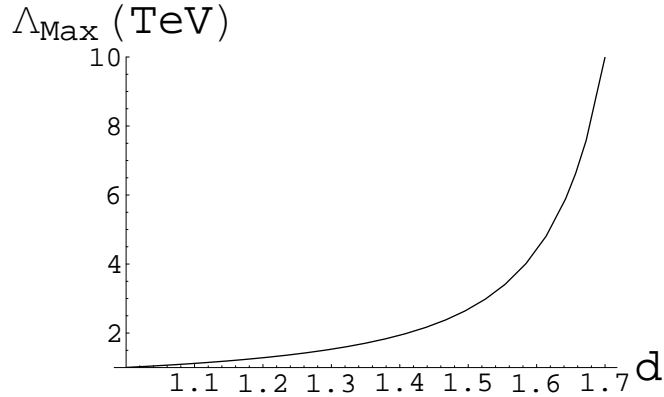}
  \caption{Qualitative behavior of $\Lambda_{max}$ as a function of $d$ for fixed $\mu$ and $\lambda_t$}
\label{fig:cutoff}
\end{figure}

The top Yukawa coupling is also important in estimating the maximum value of the cutoff allowed in the theory by using fine tuning arguments. From the usual top loop correction to the quadratic Unhiggs term in the action we find
\beq
\delta m_h^{4-2d} = \frac{3|\lambda_t|^2}{8\pi^2} \Lambda^{4-2d}~.
\eeq
Qualitatively, we want the correction to the Unhiggs mass term to be at most of the order of the tree level term to avoid excessive fine-tuning. This means we want
\beq
\label{eqn:cutoffmax}
\delta m_h^{4-2d} \approx  \frac{3|\lambda_t|^2}{8\pi^2} \Lambda^{4-2d} < \mu^{4-2d}~.
\eeq
Thus, Eq. (\ref{eqn:cutoffmax}) leads to a larger value of the maximum cutoff, $\Lambda_{max}$, for larger values of $d$, as in Figure \ref{fig:cutoff}.

In this plot, we have chosen a fixed value of $\mu$ and $\lambda_t$ consistent with the bounds required by Figure \ref{fig:TopYukCorr} at $d = 1.7$. We have also normalized so that the maximum cutoff in the Standard Model ($d=1$) is 1 TeV.
The plot clearly shows that we can push the UV scale past the usual SM value of $\sim$ 1 TeV for values of $d$ greater than 1. For example, the cutoff can be near 10 TeV without much fine-tuning for $d \sim 1.7$.

\section{Loop Induced Kinetic Term}
\label{sec:Loop}
\setcounter{equation}{0}

As we mentioned in Section 2, loop effects will also induce terms in the Lagrangian of the form 
\beq
{\cal L}_{kin} = -\frac{C}{\Lambda^{2d-2}} H^\dagger D^2 H
\label{eqn:loopp^2}
\eeq
where $C$ is a dimensionless coefficient.
Qualitatively, our analysis above is not affected by this term. However, we can estimate its quantitative effect by comparing it with the kinetic term in the original Lagrangian (\ref{action}). The ratio, $R$, of the momentum scales between the two terms is:
\beq
R = \frac{\frac{C}{\Lambda^{2d-2}}p^2}{p^{2(2-d)}} = C\left(\frac{p^2}{\Lambda^2}\right)^{d-1}~.
\eeq

Since we are considering values for an Unhiggs threshold around $\approx 100$ GeV, we take $p \approx 100$ GeV. Inserting our previous value of $\Lambda = 10$ TeV, we find
\beq
R = C(.0001)^{d-1}~.
\eeq
We expect $C < 1$ since it is a loop suppressed coefficient. For values of $d$ near one, $R \approx C$ and the loop induced term will have a relatively small quantitative effect. However, for moderate values of $d$, $R$ becomes extremely small and the term in Eq. (\ref{eqn:loopp^2}) will have no appreciable effect on the results of the previous sections.  This loop induced term will affect the region near $d=2$ where a pure unparticle is highly gaugephobic \cite{AdSCFTUnP} since it provides an additional contribution to gauge couplings.

However, since the model should be valid up to at least a few TeV, it is important to show that the loop induced kinetic term does not qualitatively change the longitudinal $WW$ scattering analysis of Section \ref{sec:production}, even for large $p^2$. After expanding Eq. (\ref{eqn:loopp^2}) around the VEV, the loop induced kinetic term contributes to the Lagrangian three terms relevant for $WW$ scattering. 
\beq
\label{eqn:LagIndKin}
\mathcal{L} \ni \frac{C}{\Lambda^{2d-2}}\left(\frac{g^2}{4}W^+W^-v^{2d} + \frac{g^2}{2}W^+W^-hv^d + \frac{1}{2}\partial_{\mu}h \partial^{\mu}h\right)~.
\eeq
This causes a modification to the $W$ mass, which is now given by
\beq
\label{eqn:ModWMass}
M_W^2 = \frac{g^2(2-d)\mu^{2-2d}v^{2d}}{4} + \frac{C}{\Lambda^{2d-2}}\frac{g^2v^{2d}}{4}~.
\eeq
The Unhiggs propagator and the $hW^+W^-$ vertex will also be modified by Eq. (\ref{eqn:LagIndKin}). The pure gauge contribution (Fig. \ref{fig:WWSM}) to the $WW$ scattering amplitude will retain the same form as in Eq. (\ref{eqn:WWSM}), but with the modified $W$ mass. Also, the four gauge boson two Unhiggs contribution (Fig. \ref{fig:WW2UnH}) is still given by Eq. (\ref{eqn:2Unhiggs s}), again with the modified $W$ mass. The contribution from the two gauge boson one Unhiggs vertex (Fig. \ref{fig:WW1UnH}) contains both the modified $hW^+W^-$ vertex as well as the modified Unhiggs propagator. The s-channel contribution to the longitudinal $WW$ scattering amplitude is now given by
\beq
\mathcal{M}_{1h} = \frac{-ig^4v^{2d}}{4}\left(1-\frac{s}{2M_W^2}\right)\frac{\left(\frac{C}{\Lambda^{2d-2}}+\frac{\mu^{4-2d}-(\mu^2-s)^{2-d}}{s}\right)^2}{\mu^{4-2d}-m^{4-2d}-(\mu^2-s)^{2-d}+\frac{C}{\Lambda^{2d-2}}s}~.
\eeq
Upon taking $s \gg M_W,\mu,m$, we get
\beq
\label{eqn:IndKinWW1Unh}
\mathcal{M}_{1h} = \frac{-ig^4v^{2d}}{16M_W^4}\left[\frac{C}{\Lambda^{2d-2}}s-(-s)^{2-d}\right]~.
\eeq
Combining the contributions from Eqs. (\ref{eqn:WWSM}), (\ref{eqn:2Unhiggs s}) and (\ref{eqn:IndKinWW1Unh}) yields the following for the s-channel contribution to the $WW$ scattering amplitdue:
\begin{align}
\mathcal{M}_{s-channel} &= \frac{-ig^4v^{2d}}{16M_W^4}\frac{C}{\Lambda^{2d-2}}s + \frac{ig^2}{4M_W^2}\left[1-\frac{g^2v^{2d}}{4M_W^2}(2-d)\mu^{2-2d}\right]s \\ \nonumber
& = \frac{ig^2}{4M_W^2}\left[1-\frac{g^2v^{2d}}{4M_W^2}\left(\frac{C}{\Lambda^{2d-2}}+(2-d)\mu^{2-2d}\right)\right]s~.
\end{align}
Using the modified formula for the $W$ mass in Eq. (\ref{eqn:ModWMass}), we find
\begin{align}
\mathcal{M}_{s-channel} &= \frac{ig^2}{4M_W^2}\left(1-\frac{g^2v^{2d}}{4M_W^2}\frac{4M_W^2}{g^2v^{2d}}\right)s \\ \nonumber
& = \frac{ig^2}{4M_W^2}(1-1)s = 0~.
\end{align}
The analysis of the t-channel contribution is exactly analogous. Thus, we have shown that for $s,t \rightarrow \infty$, the loop induced kinetic term does not affect the unitarity of longitudinal $WW$ scattering. Of course, the loop induced kinetic term will affect the quantitative analysis of the allowed values of the parameters in Figure \ref{fig:Contourplot}, but since we expect C to be loop suppressed, the loop induced kinetic term should not drastically affect the results.


\section{Conclusions}

We have explored the possibility that the Higgs arises from an approximately conformal sector and is described by an unparticle. We have found that such an Unhiggs can still break electroweak symmetry and unitarize $WW$ scattering, but that the lower bounds  on the mass threshold from LEP are much weaker than for a SM Higgs.  Raising the scaling dimension of the Unhiggs mass term serves to weaken the little hierarchy problem since the power dependence on the cutoff is reduced.  This is as one would expect, since breaking electroweak symmetry by an operator with dimension greater than two and thus a mass term operator dimension greater than four (at least at weak coupling) provides a solution to the full hierarchy problem. This is essentially what happens in  Randall-Sundrum models (and other technicolor-like models): the scaling dimension of the (analogue) of the Higgs mass operator is very large and thus safe from any divergences.  For an Unhiggs the mass term operator dimension is between two and four, an thus the mass term divergence scales as the cutoff to a power between two and zero. It would also be very interesting to explore how Unhiggs loops affect precision electroweak measurements.

\section*{Acknowledgments}
We thank Markus Luty, Matt Strassler, Giacomo Cacciapaglia, Csaba Cs\'aki, Hsin-Chia Cheng, Jamison Galloway, John McRaven and Damien Martin for useful discussions and comments.
The authors are supported by the US department of Energy under contract DE-FG02-91ER406746.

\appendix

\end{document}